\documentclass[superscriptaddress,twocolumn]{revtex4}

\usepackage{graphicx}
\usepackage{epsfig}
\usepackage{dcolumn}
\usepackage{bm}
\usepackage{amsmath}
\usepackage{amssymb}
\usepackage{latexsym}
\usepackage[usenames,dvipsnames]{xcolor}
\usepackage{color}

\newcommand\B{\rule[-1.2ex]{0pt}{0pt}}
\newcommand\T{\rule{0pt}{2.6ex}}

\begin{document}

\title{Dimensional fragility of the Kardar--Parisi--Zhang universality class}

\author{Matteo Nicoli} 
\address{Center for Interdisciplinary Research on Complex Systems,
Department of Physics, Northeastern University, Boston, MA 02115, USA.}

\author{Rodolfo Cuerno}
\address{Departamento de
Matem\'{a}ticas and Grupo Interdisciplinar de Sistemas Complejos
(GISC), Universidad Carlos III de Madrid, Avenida de la
Universidad 30, E-28911 Legan\'{e}s, Spain.}

\author{Mario Castro}
\address{GISC and Grupo de Din\'amica No Lineal (DNL), Escuela T\'ecnica
Superior de Ingenier{\'\i}a (ICAI),  Universidad Pontificia Comillas, E-28015
Madrid, Spain.}

\begin{abstract}
We assess the dependence on substrate dimensionality of the asymptotic scaling behavior of a whole family of equations
that feature the basic symmetries of the Kardar--Parisi--Zhang (KPZ) equation. Even for cases in which, as expected from universality arguments, these models display KPZ values for the critical exponents and limit distributions, their behavior deviates from KPZ scaling for increasing system dimensions. Such a fragility of KPZ universality contradicts naive expectations, and questions straightforward application of universality principles for the continuum description of experimental systems.
\end{abstract}

\maketitle

One of the most powerful concepts in contemporary Statistical Mechanics is the idea of universality,
by which microscopically dissimilar systems show the same large scale behavior, provided
they are controlled by interactions that share dimensionality, symmetries, and conservation laws. Being rooted in the behavior of equilibrium critical systems \cite{eq_review}, universality has more recently allowed to describe scaling behavior far from equilibrium \cite{marro:book,odor:2004,munoz:2011}, as for e.g.\ the stock market \cite{meyers:2011}, crackling--noise \cite{sethna:2001}, or random networks \cite{albert:2002}. In complex systems like these, universality provides an enormously simplifying framework, as significant descriptions can be put forward on the basis of the general principles just mentioned.

Celebrated non-equilibrium systems include those with generic scale invariance, displaying criticality throughout parameter space \cite{belitz:2005}. Examples are
self--organized--critical \cite{grinstein:1995} and driven--diffusive systems \cite{schmittmann:2000}, or surface kinetic roughening \cite{krug:1997}. Indeed, the paradigmatic Kardar--Parisi--Zhang (KPZ) equation for a rough interface \cite{kardar:1986}
\begin{equation}
\partial_t h = \nu \nabla^2 h + \frac{\lambda}{2} (\nabla h)^2 + \eta(\mathbf{x},t) ,
\label{kpz}
\end{equation}
is very recently proving itself as a remarkable instance of universality. Here, $h(\mathbf{x},t)$ is a height field above substrate position $\mathbf{x} \in \mathbb{R}^d$ at time $t$, and $\eta$ is Gaussian white noise with zero mean and variance $2D$. Thus, the exact asymptotic height distribution function has been very recently obtained for $d=1$ \cite{sasamoto:2010,amir:2011,calabrese:2011}: it is given by the largest--eigenvalue distribution of large random matrices in the Gaussian unitary (GUE) (orthogonal, GOE) ensemble, the Tracy--Widom (TW) distribution, for globally curved (flat) interfaces, as proposed in \cite{praehofer:2000}, see reviews in \cite{kriecherbauer:2010,corwin:2012}. Besides elucidating fascinating connections with probabilistic and exactly solvable systems, these results are showing that, not only are the critical exponent values common to members of this universality class, but also the distribution functions and limiting processes are shared by discrete models and continuum equations \cite{alves:2011_etal}, and by experimental systems, from turbulent liquid crystals \cite{takeuchi_et_al} to drying colloidal suspensions \cite{yunker:2013}.

In view of the success for $d=1$ (1D) substrates, a natural important step is to assess the behavior of the KPZ universality class when changing space dimension, analogous to e.g.\ the experimental change from 2D to 1D behavior for ferromagnetic nanowires, that nonetheless occurs within the creeping--domain--wall class \cite{kim:2009}. Thus, for discrete models and the continuous equation itself, indeed universal distributions have been also very recently found \cite{halpin-healy:2012,oliveira:2013} to control height fluctuations in the $d=2$ KPZ class, providing analogs of the TW distributions. Again this underscores universality, beyond critical exponents already known to be shared by the KPZ equation, many discrete models \cite{kelling:2011}, and some experiments \cite{eklund:1991_et_al}, although much less than expected \cite{cuerno:2007}. This fact calls for further experimental verification, akin to that recently provided \cite{takeuchi_et_al,yunker:2013} for the one--dimensional case.

In this work we report a fragility of the KPZ universality class with respect to space dimension. Namely,
we consider a family of continuum equations with the symmetries and conservation laws of the KPZ equation.
We provide conditions under which, although the system is accurately described by KPZ universality in $d=1$, the scaling exponents {\em depart} from the latter in $d=2$. Analogous behavior had been found earlier for discrete models of conservative surface growth \cite{dassarma:2002}, namely, a change of the universality class of a given system with $d$. Here we demonstrate it for non--conserved dynamics, and at the level of continuum equations. Note, this is not a change in the universality class as a response to changes in appropriate system parameters for a fixed dimension, as seen e.g.\ for Barkhausen criticality \cite{ryu:2007}. The lack of universality that we find signals a serious difficulty in the identification of the appropriate universality class for experimental systems, by preventing cursory use of universality arguments to propose theoretical descriptions, stressing the need for physically motivated models \cite{cuerno:2007}. This should be borne in mind, in view of the timely interest in the experimental validation of 2D KPZ universality.

\begin{figure*}[!t]
\centering
\includegraphics[angle=0,width=0.75\textwidth]{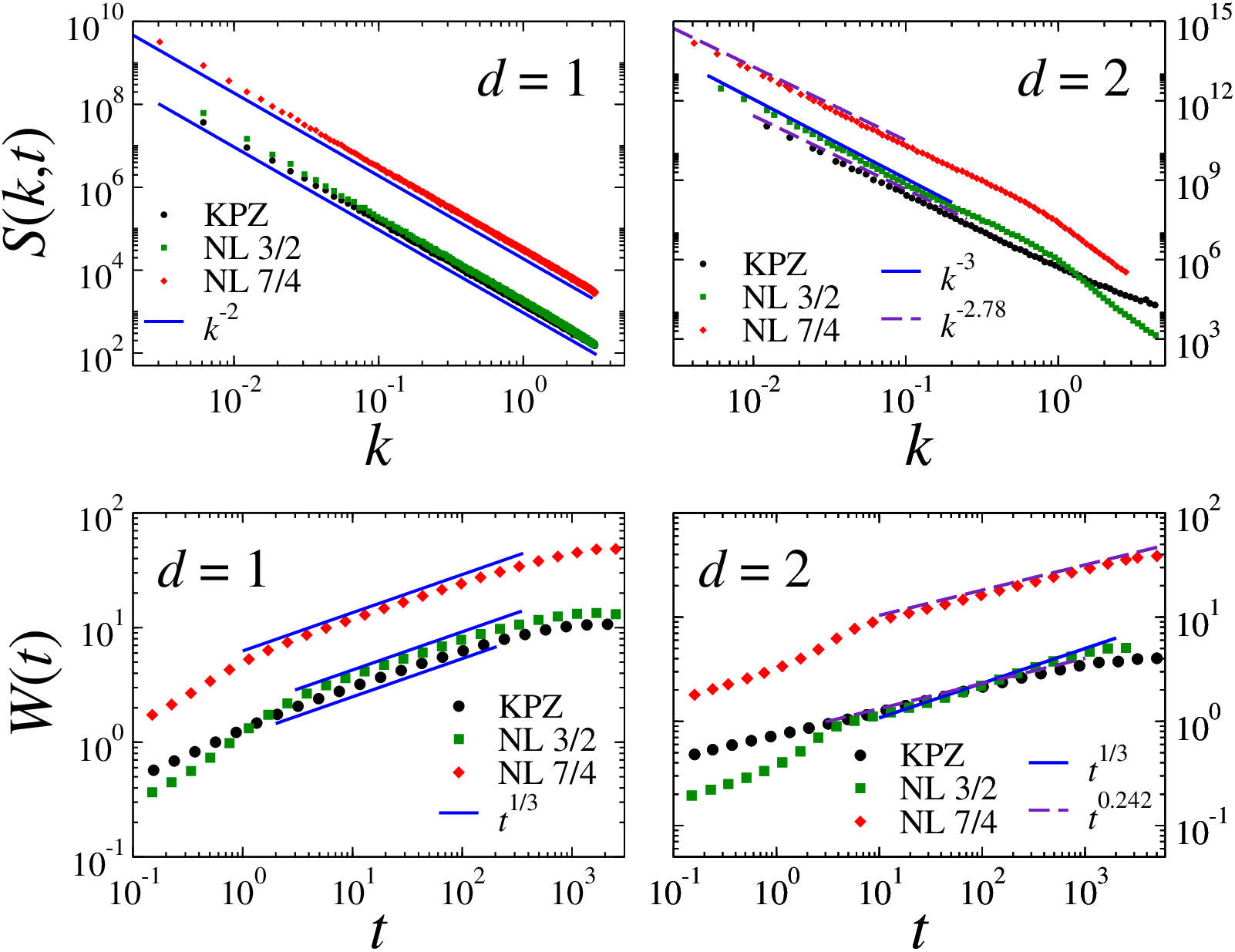}
\caption{ Numerical simulations of Eq.\ \ref{kpz} ($\bullet$) and Eq.\ \ref{fake} for $\mu=3/2$ ({\color{OliveGreen}$\scriptstyle{\blacksquare}$}) and $\mu=7/4$ ({\color{red}$\blacklozenge$}).
Surface structure factor (upper row) and roughness (lower row) for $d=1$ (left column) and $d=2$ (right column). Solid and dashed  lines represent power-law behaviors as indicated, which are discussed in the main text. All the observables have been averaged over $10^3$ ($10^2$) different noise realizations for $d=1$ ($d=2$), starting from a flat initial condition. Error bars are smaller than symbol sizes. All units are arbitrary.}
\label{fig:1}
\end{figure*}

We consider the following equation \cite{nicoli:2009}
\begin{equation}
\partial_t h_{\mathbf{k}}(t) = (\nu k^{\mu} - {\cal K} k^2) h_{\mathbf{k}}(t) + \frac{\lambda}{2} {\cal F}[(\nabla h)^2]_{\mathbf{k}} + \eta_{\mathbf{k}}(t) ,
\label{fake}
\end{equation}
where $\nu, {\cal K} >0$, $k = |\mathbf{k}|$, and ${\cal F}$ is space Fourier transform, with $h_{\mathbf{k}}(t)$ and $\eta_{\mathbf{k}}(t)$ being the $\mathbf{k}$-th modes of the height and noise fields, respectively. Equation \ref{fake} perturbs the KPZ equation \ref{kpz} through the linear term with coefficient $\nu$, where $0 < \mu < 2$. This family of equations includes celebrated systems, such as the Kuramoto--Sivashinsky (KS) (take $\mu\to 2$ and replace $k^2$ with $k^4$) and the Michelson--Sivashinsky ($\mu=1$) equations \cite{nicoli:2009}, that combine pattern formation at short scales with asymptotic kinetic roughening \cite{misbah:2010}. For specific $\mu$ values, Eq.\ \ref{fake} describes {\em quantitatively} experiments of diffusion--limited interface processes, like plasma etching \cite{zhao:1999}, electrochemical \cite{nicoli:2009b}, and chemical vapor \cite{castro:2012} deposition.

\begin{table}[t!]
\centering
\begin{tabular}{ll|ccccccc}
\hline
\hline
\multicolumn{2}{c|}{Equation} & $\nu$ & $\mathcal{K}$ & $\lambda$ &$D$ & $L$ & $\Delta x$ & $\Delta t$  \T  \B\\
\hline
 		& KPZ & 1  & & 5.0 & 1.46 & 1024 & 1.0 & $0.002$  \T  \\
$d=1$ 	& NL $3/2$ & 1  & 1.0    & 5.0 & 0.50 & 1024 & 1.0 & $0.001$  \T  \\
		& NL $7/4$ & 1 & 1.7 & 2.5 & 12.5 & 2048 & 2.0 & $0.002$  \T  \\
\hline
		& KPZ & 1 & & 2.5 & 1.46 & 512 & 1.0 & $0.01$  \T \\
$d=2$	& NL $3/2$ & 1  & 0.5 & 10 & 0.50 & 1024 & 1.0 & $0.01$  \T \\
		& NL $7/4$  & 1 & 1.0 & 1.0& 50.0 & 1536 & 1.5 & $0.02$  \T  \\
\hline
\hline
\end{tabular}
\caption{Parameters used for the numerical integrations reported in this work. NL stands for the non-local models,
i.e.\ Eq.\ \ref{fake} with $\mu$ equal to $3/2$ or $7/4$. $L$ is the size of the 1D domain, or the edge of the 2D square, used for simulations in Fig.\ \ref{fig:1}.
All units are arbitrary.}
\label{tab_fig1}
\end{table}

\begin{table}[b!]
\centering
\begin{tabular}{ll|ccccccc}
\hline
\hline
\multicolumn{2}{c|}{Equation} & $v_\infty$ & $\Gamma$ & $t^*$ &  $c_v^{\rm est}$ & $c_v$ 	& Abs. error \T  \B\\
\hline
 		& KPZ  & 3.6990   & 5.620 & 1000 & 0.45		& 0.45045&	0.01\% \T  \\
$d=1$ 	& NL $3/2$  & 4.2675   & 11.95 & 500  & 0.52		& 0.57923&	10\% \T  \\
		& NL $7/4$  & 18.4625 & 310.0 & 500 & 1.56		& 1.71469&	9\% \T  \\
\hline
\hline
\end{tabular}
\caption{Values of the constants used for the determination of the the probability distribution function $P(\chi)$.
Here, $t^*$ is the time used in the computation of the quantities reported in Figs.\ \ref{fig:2} and \ref{fig:3}.
In the last column we report the absolute error between  calculated coefficient $c_v =  \beta \Gamma^\beta \langle \chi \rangle$ and its estimated value $c_v^{\rm est}$ from our numerical data
for the three equations in case of one-dimensional substrates ($d=1$).}
\label{tab_fig2}
\end{table}

Equation \ref{fake} complies with the standard symmetries of the KPZ class. Namely, dynamics are non--conserved, isotropic and reflection invariant in $\mathbf{x}$, invariant under Galilean transformations and under arbitrary shifts $h \to h +{\rm const.}$, and the up--down symmetry $h \leftrightarrow -h$ is broken \cite{nicoli:2011}. Two additional features are to be noted, namely, the non--analytic dependence \cite{kassner:2002} on $\mathbf{k}$, and the morphological instability \cite{cuerno:2007}. The former induces {\em non--locality} of the equation when written in real space \footnote{The real-space representation of $k^{\mu} h_{\mathbf{k}}$ is proportional \cite{nicoli:2011} to the Cauchy principal value of $\int_{\mathbb{R}^d} [h(\mathbf{r})-h(\mathbf{r}')]/|\mathbf{r}-\mathbf{r}'|^{d+\mu} \, {\rm d}\mathbf{r}'$ for $0 < \mu \leq 2$.} \footnote{Eq.\ \ref{fake} is a non-equilibrium system with weakly long-range interactions, see \cite{mukamel:2010}.}. The latter breaks scale invariance at short time and length scales, which is restored back at large scales along the dynamics, as in the KS system \cite{krug:1997}. Indeed, as borne out by numerical \cite{nicoli:2009} and dynamic renormalization group \cite{nicoli:2011} results, the asymptotic behavior of Eq.\  \ref{fake} fulfills the Family-Vicsek (FV) scaling ansatz \cite{krug:1997}. Hence, the surface structure factor, $S(k,t)= \langle |h_{\mathbf{k}}(t)|^2 \rangle$, scales at long times as $S(k,t\to\infty) \sim 1/k^{2\alpha+d}$, with a well--defined value of the roughness exponent $\alpha$. The crossover wave--vector value separating white noise from correlated behavior also scales, $k_c \sim t^{-1/z}$, leading to power--law behavior of the global roughness $W(t)$ (root mean square fluctuation of the surface height) with time as $W \sim t^{\beta}$, with $\beta=\alpha/z$. The values of these critical exponents depend on $\mu$, and correspond to the KPZ universality class in $d$ dimensions, provided $z_{\rm KPZ}(d) \leq \mu < 2$, where $z_{\rm KPZ}(d)$ is the corresponding KPZ value of the dynamic exponent. For $\mu< z_{\rm KPZ}(d)$ and any $d$, the asymptotic exponents are non-KPZ, namely, $z=\mu$ and $\alpha=2-z$ \cite{nicoli:2009}. Note that, for the morphologically unstable condition $\nu>0$ that we consider, the nonlinearity is dynamically relevant for any value of $\mu$ (even if it may not control scaling behavior), while for the morphologically stable situation ($\nu < 0$) scaling in Eq.\ \ref{fake} is controlled by the linear terms for small $\mu$ values \cite{nicoli:2009,nicoli:2011,katzav:2003}.

In the left column of Fig.\ \ref{fig:1} we show 1D numerical simulations of Eq.\ \ref{fake} for $\mu=3/2,7/4$ (denoted as NL, for non--local) and, as a reference, for the KPZ equation itself, using a pseudospectral scheme as in \cite{nicoli:2009,nicoli:2009b} and parameters reported in Table \ref{tab_fig1}. Both values of $\mu\geq z_{\rm KPZ}(1)=3/2$, thus for $d=1$ the behavior is well described by KPZ exponents \mbox{$\beta_{\rm KPZ}(1)=1/3$} and $\alpha_{\rm KPZ}(1)=1/2$. 1D simulations with non--KPZ exponents for $\mu < z_{\rm KPZ}(d)$ can be found in \cite{nicoli:2009}.

\begin{figure*}[!t]
\centering
\includegraphics[angle=0,width=0.9\textwidth]{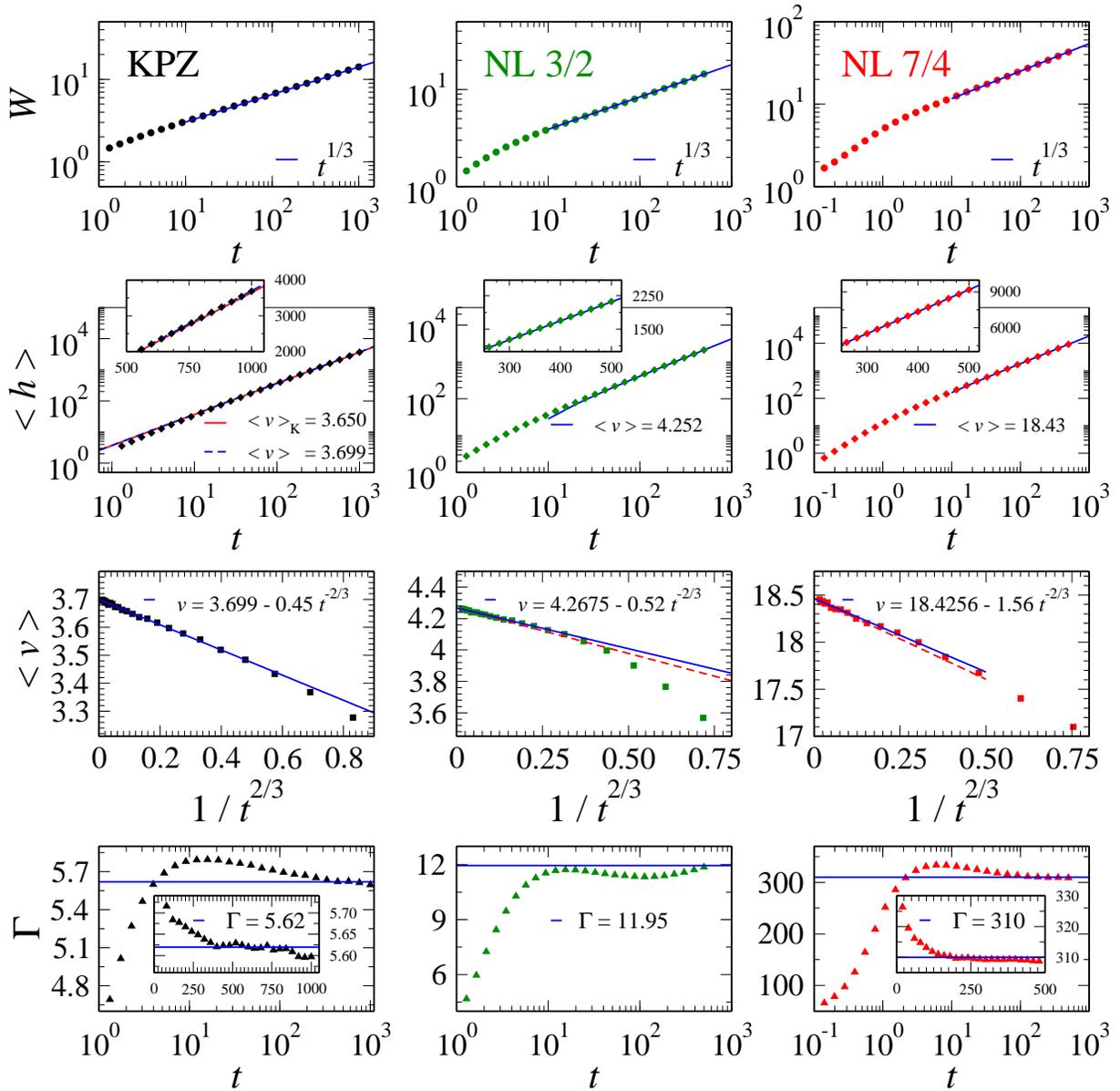}
\caption{Estimation of $v_\infty$ and $\Gamma$ for the one--dimensional KPZ and Non--Local ($\mu=3/2,7/4$) equations. For the former, the value of $v_\infty$ is equally well measured from $\langle \bar{h} \rangle$ and $\langle \bar{v} \rangle$ data, the red solid line in the uppermost--right panel being the the value of $v_\infty$ in the limit of an infinite system size $\langle \bar{v} \rangle_{\rm K} = D\lambda / 2 \Delta x$ (with $\Delta x = 1$), calculated in \cite{krug:1997} (note that our noise variance is $2D$ and not $D$). For the NL model, the value of $v_\infty$ differs if considered from $\langle \bar{h} \rangle$ or from $\langle \bar{v} \rangle$ data. Here we take the latter choice because the $\langle \bar{v} \rangle$ data are smooth and do not display large fluctuations. The red dashed lines in the corresponding $\langle \bar{v} \rangle$ vs $1/t^{2/3}$ plots implement the consistency fit as calculated from the parameters reported in Table 1.}
\label{fig_estimation}
\end{figure*}

KPZ universality here {\em goes beyond} exponent values. Thus, using the Ansatz \cite{praehofer:2000}
\begin{equation}
h(x,t) \simeq v_{\infty} t + {\rm sgn}(\lambda) (\Gamma t)^\beta \chi,
\label{eq_h}
\end{equation}
we can measure the fluctuations of the interface around its mean value  $v_{\infty}t$, i.e.
\mbox{$\chi = {\rm sgn}(\lambda) (h-v_\infty t) / (\Gamma t)^\beta$}. Through $\Gamma$ we normalize
the variance of this stochastic process to the variance of the TW--GOE distribution  and
we are able to compare them. For the estimation of $v_\infty$ and $\Gamma$ we followed the procedure described in \cite{alves:2011_etal}. Values for these constants are reported in Table \ref{tab_fig2}. Specifically, $\Gamma$ and $v_\infty$ are calculated by averaging quantities measured on interface profiles $h(x,t)$. Thus, after time differentiation of Eq.\ \ref{eq_h}, we obtain an Ansatz for the instantaneous velocity of each point of the interface,
\begin{equation}
v(x,t) \simeq v_\infty + \beta\, {\rm sgn}(\lambda) \Gamma^\beta t^{\beta-1} \, \chi,
\end{equation}
so that the average of this observable reads
\begin{equation}
\label{mean_v}
\langle \bar{v} \rangle = \frac{d}{dt}\langle \bar{h} \rangle = v_\infty + \beta\, {\rm sgn}(\lambda) \Gamma^\beta t^{\beta-1} \,
	\langle \chi \rangle,
\end{equation}
provided $\langle \chi \rangle \neq 0$, where the overline stands for spatial averages on the same interface, while
brackets refer to average over different runs. Plots $\langle \bar{h}\rangle$ vs $t$, or $\langle \bar{v}\rangle$ vs $t^{\beta-1}$ are used to measure $v_\infty$, depending on each equation we consider. Comparison of $P(\chi)$ with the TW--GOE distribution is possible only after we estimate $v_\infty$ very accurately (up to the fourth decimal place); small errors result into a misalignment of the maxima of the two distributions.
The second step is to normalize the variance of $\chi$ to the variance of the TW--GOE distribution by using the parameter
$\Gamma$. From the variance of $h$ (that is, the second cumulant $\langle h^2 \rangle_c$) we cancel the $v_\infty t$ contribution so that $\langle h^2 \rangle_c \simeq \left(\Gamma t \right)^{2\beta} \langle \chi^2\rangle_c$, and, by normalizing the variance of $\chi$ to \mbox{$\langle \chi^2_{\small \rm GOE}\rangle_c = 0.638$}, we get
\begin{equation}
\Gamma = t^{-1} \left(\frac{\langle h^2\rangle_c}{\langle \chi^2_{\rm GOE}\rangle_c} \right)^{1/2\beta}.
\end{equation}
We show in Fig.\ \ref{fig_estimation} the implementation of this procedure to estimate $v_\infty$ and $\Gamma$ for the one-dimensional KPZ and Non-Local ($\mu=3/2,7/4$) equations, leading to the values quoted in Table \ref{tab_fig2}. Moreover,  we perform a consistency check of Eq. \ref{mean_v} through the independent estimation of $v_\infty$ and $\Gamma$. Namely,  the $\Gamma$-dependent quantity $c_v(\Gamma) = \beta \Gamma^\beta \langle \chi \rangle$ calculated from $\Gamma$ must match the coefficient of the time behavior of $\langle v \rangle$ obtained from numerical data. This calculation is straightforward for $d = 1$ equations where the limit distribution $\chi$ is known and $\langle \chi \rangle = -0.76007$, see Table \ref{tab_fig2}. For the KPZ equation the agreement is remarkable, but for nonlocal equations errors are
not so small. For these equations, the transient regime between the initial exponential growth due to the  pattern formation process and the asymptotic scale invariant state hinders a more clear match; probably, longer times are required in order to get a better agreement between  $c_v(\Gamma)$ and the coefficient from $\langle v\rangle$ data.
Still, the red dashed lines plotted in Fig.\ \ref{fig_estimation} show that even a fit with the calculated coefficient $c_v$ (with $\sim 10\%$ absolute error) is compatible with our numerical data.

\begin{figure}[!t]
\centering
\includegraphics[angle=0,width=0.5\textwidth]{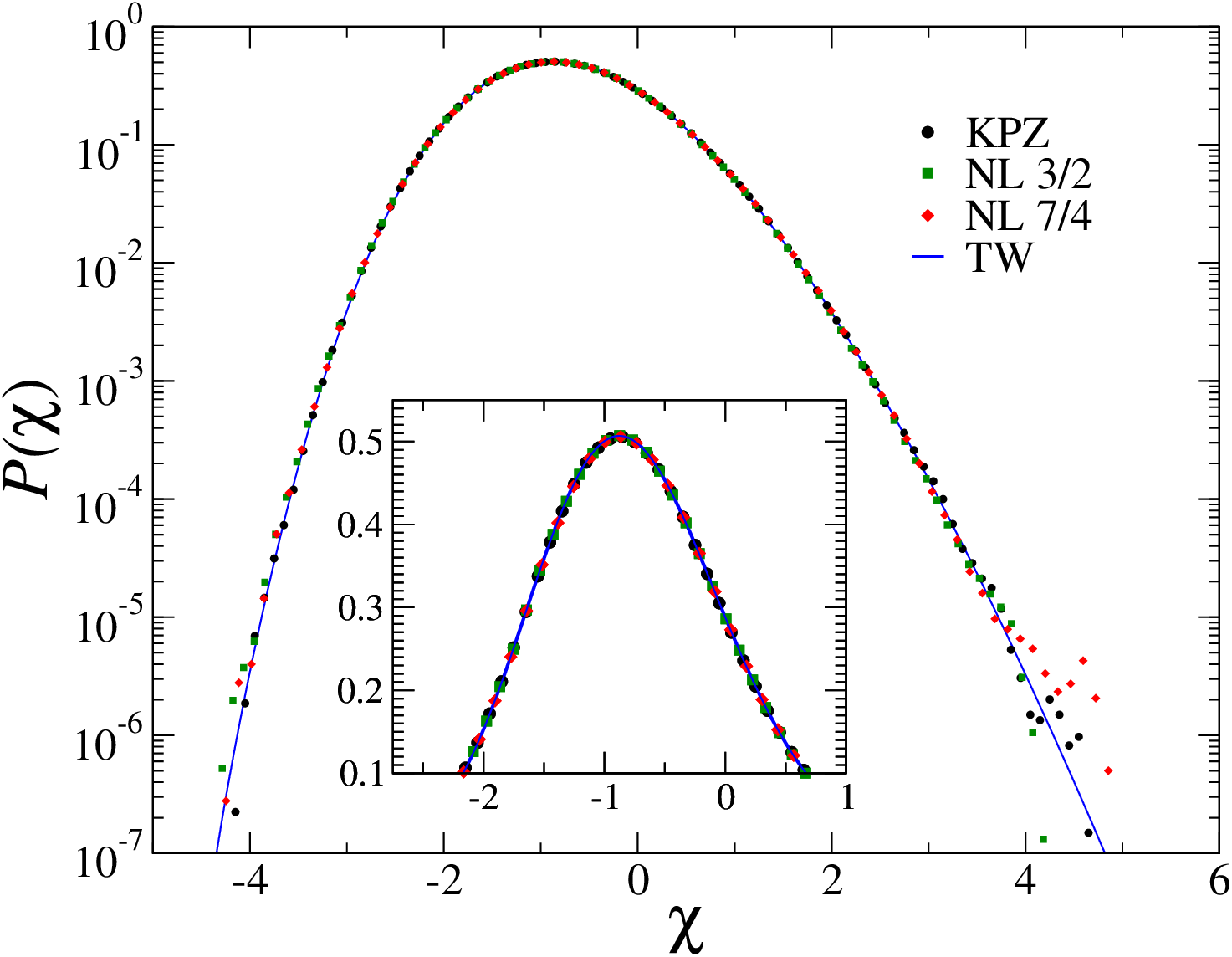}
\caption{1D Height distributions for Eq.\ \ref{kpz} ($\bullet$) and Eq.\ \ref{fake} for $\mu=3/2$
({\color{OliveGreen}$\scriptstyle{\blacksquare}$})  and $\mu=7/4$ ({\color{red}$\blacklozenge$}). The variable $\chi$ is defined in the text. The solid blue line is the TW (GOE) distribution expected for $d=1$ \cite{edelman:2008}. $P(\chi)$ is estimated from 2048 independent runs starting from a flat initial condition. Inset: zoom of main panel, in linear representation. All units are arbitrary.}
\label{fig:2}
\end{figure}

\begin{figure}[!t]
\centering
\includegraphics[angle=0,width=0.5\textwidth]{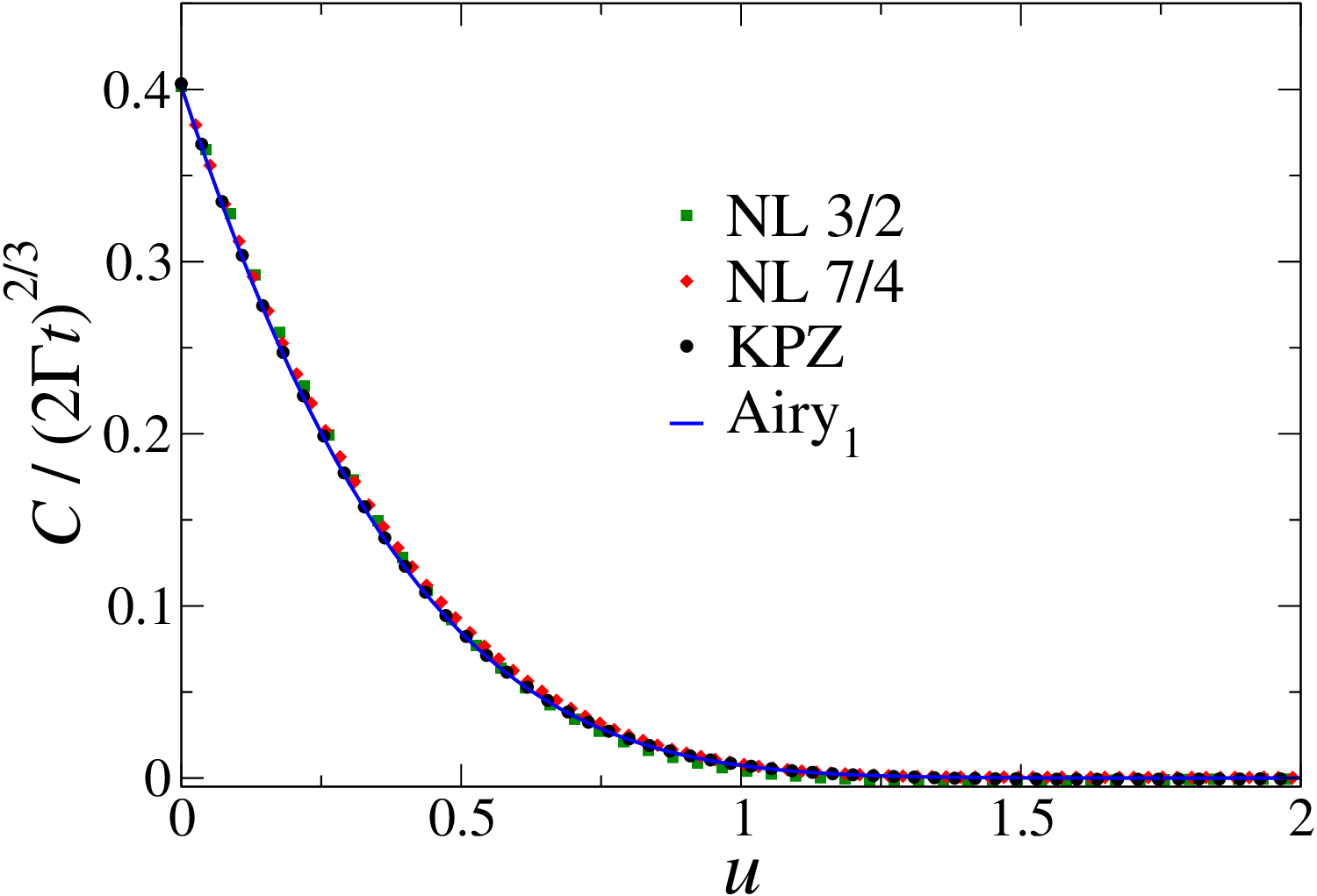}
\caption{ 1D height-height correlation function for Eq.\ \ref{kpz} ($\bullet$) and Eq.\ \ref{fake} for $\mu=3/2$ ({\color{OliveGreen}$\scriptstyle{\blacksquare}$}) and $\mu=7/4$ ({\color{red}$\blacklozenge$}). Here,
$u= x\sqrt{\Gamma/2\lambda}/(2\Gamma t^*)^{2/3}$ while $C(x,t=t^*)$ is measured from the same surfaces employed to estimate $P(\chi)$ in Fig.\ \ref{fig:2}. The solid blue line provides the covariance of the Airy$_1$ process \cite{bornemann:2010}. All units are arbitrary.}
\label{fig:3}
\end{figure}

Once $v_\infty$ and $\Gamma$ have been determined, we can assess the probability distribution for the rescaled height fluctuations $\chi$ in each one of the equations we are studying. Numerical results are provided in Fig.\ \ref{fig:2}, where they are compared with the analytical GOE TW distribution, shown as a solid line. The latter has been calculated from the solution of the Painlev\'e II differential equation \cite{edelman:2008}, and normalized according to \cite{praehofer:2000}.
Clearly, the $P(\chi)$ distribution is time-independent and indeed agrees with the analytical result for the three equations, providing in particular an assessment of the TW distribution also for non-local equations.
Further universal behavior is seen to occur, as in other instances of the 1D KPZ class \cite{alves:2011_etal}, for the two--point correlation \mbox{$C(x,t) = \langle h(x_0+x,t)h(x_0,t)\rangle - \langle h\rangle^2$}, that scales as $C(x,t) \sim (2\Gamma t)^{2/3} g_1(u)$, where $u=(A x/2)/(2\Gamma t)^{2/3}$ with $g_1(u)$ the covariance of the Airy$_1$ process \cite{alves:2011_etal,bornemann:2010}, and $A = (2\Gamma/\lambda)^{1/2}$ for continuum equations (for discrete models, $A$ is estimated from the local roughness \cite{alves:2011_etal}). As suggested by \mbox{Fig.\ \ref{fig:2}}, even for the 1D non--local models the rescaled two--point correlation function collapses perfectly onto $g_1(u)$, see Fig.\ \ref{fig:3}.

Hence, in $d=1$ all the strong universal properties of the KPZ class occur in the NL $\mu=3/2$ and $\mu=7/4$ systems.
However, when we increase the system dimension to $d=2$, a remarkable departure from KPZ scaling occurs that depends on the relative values of $\mu$ and \mbox{$z_{\rm KPZ}(2)\approx 1.61$}. Thus, while Eq.\ \ref{fake} is still well described by KPZ exponent values for ``large'' $\mu=7/4>1.61$ as deduced from Fig.\ \ref{fig:1}, namely, $\alpha\simeq 0.39$
(compare $\alpha_{\rm KPZ}(2)\approx 0.39$ \cite{kelling:2011}) and $z\simeq 1.61$, the ``small'' $\mu=3/2<1.61$ system has the same exponent values as for $d=1$! Recall that, for $z_{\rm KPZ}(1) \leq \mu<z_{\rm KPZ}(2)$, the 2D exponents are non--KPZ \cite{nicoli:2009}, $z=\mu$, $\alpha=2-z$, moreover they are $d$--independent. Curiously enough, thus the $\mu=3/2$ equation provides a peculiar example of a 2D system with 1D-KPZ exponents! Without the need of further characterization of height distributions or correlation functions, for this equation this implies a change of its universality class as dimensionality increases from $d=1$ to $d=2$, while this is not the case for e.g.\ the $\mu=7/4$ equation, which is still KPZ-like in 2D.

This fact bears important consequences on the continuum modeling of systems, in particular of an experimental type, that are presumably in the KPZ universality class.
Take the 1D case as an example. Eq.\ \ref{fake} having the same symmetries as the KPZ equation, one might postulate the latter as a model description for a given experiment. But suppose the actual physical interactions lead to the occurrence of morphological instabilities (as it happens only too often in surface growth experiments \cite{cuerno:2007}), in such a way that a better description is provided by Eq.\ \ref{fake} for $\mu=3/2$. This will not change the 1D scaling behavior with respect to KPZ universality, even at the level of height distributions or correlation functions. However, if one were able to perform an experiment for the 2D generalization of the system, a departure from KPZ behavior would be obtained, with the conclusion that the universality class of the physical system would {\em not} be KPZ. One might argue that increasing $d$ for a fixed $\mu$ makes interactions more non--local in real space \cite{mukamel:2010}, and that the present fragility of KPZ scaling is only superficial \cite{munoz:2011}. However, this does not circumvent the need, for a given physical system, to assess in detail the occurrence of e.g.\ morphological instabilities and/or the range of interactions, in order to argue for the correct universality class on a safe basis. In any case, this requires going beyond symmetry principles to provide the sought--for continuum description. Note that, starting out with a higher value of $\mu$ that leads to KPZ scaling both in $d=1$ and 2, such as $\mu=7/4$, only pushes departure from KPZ scaling up to a higher dimension $d_{7/4}$, such that $z_{\rm KPZ}(d_{7/4}) > 7/4$, which will occur below the upper (if finite) critical dimension $d_c$ for the KPZ universality class, at which $z_{\rm KPZ}(d_c)=2$.


Summarizing, we have found a fragility of the KPZ universality class with respect to space dimension, when perturbed by morphological instabilities combined with non--local interactions, within the {\em experimentally substantiated} family of equations, Eq.\ \ref{fake}. Note, an important perturbation of the KPZ equation by instabilities that respects its space symmetries, is also known to occur in the celebrated (noisy) KS system, which is a local equation known to lead to KPZ scaling, both in $d=1$ \cite{krug:1997} and $d=2$ \cite{nicoli:2010}. We recall that earlier results have also suggested non-universal behavior for the KPZ class in $d>1$. E.g.\ for increasing $d$, details of the noise distribution have been reported to become relevant \cite{newman:1997}. Important issues remain indeed open with respect to the behavior of the KPZ class in higher dimensions, like the existence and value of an upper critical dimension (see e.g.\ \cite{canet:2010} and references therein), or even making mathematical sense of solutions to Eq.\ \ref{kpz} \cite{corwin:2012}.

Although non--equilibrium universality classes are frequently expected to be more fragile than equilibrium ones \cite{marro:book}, a highly non--trivial question is to identify, if existent, the type of perturbation that is taking place here and assess its actual importance \cite{munoz:2011}. In our case, even if the present fragility might be relativized in view of the non--local nature of the perturbation, its occurrence is not easily circumvented by the symmetry arguments that are usually in use for the theoretical description of kinetic roughening phenomena. This seems an important {\em caveat}, especially in view of the current quest for 2D KPZ scaling behavior in {\em experimental} systems. As implied by our results, in order to assign a universality class to a given system, one would need to explore its behavior under a change in $d$. However, for many experimental systems modifying the space dimension may be hard to achieve without significantly altering the basic interactions that take place. For instance, basic properties of fluid flow can drastically change from a quasi--2D Hele--Shaw cell to a 3D system, while keeping all additional conditions unchanged \cite{batchelor:book}. This stresses the need for detailed modeling of the specific peculiarities of the system under study, undermining the promise of universality as the main toolbox for kinetically rough systems. An analogous situation occurs in the context of pattern formation, where Goldstone modes associated with the shift symmetry $h\to h +{\rm const.}$ prevent the existence of a universal amplitude equation \cite{hoyle:book}. In such contexts, modeling has to be done on a system-specific basis. We note that in these cases symmetry arguments can be enhanced by multiple scales approaches in order to put forward general continuum models that successfully describe experimental systems \cite{castro:2007}. One can ponder \cite{kassner:2002} whether analogous generalized approaches would be successful in the presence of non-localities and noise.


\acknowledgments
We thank E.\ Moro, J. Rodr\'{\i}guez-Laguna,  S.\ N.\ Santalla, and P.\ Vivo for discussions, and F.\ Bornemann for providing the $g_1(u)$ function. This work has been partially supported through grants FIS2009-12964-C05-01 and FIS2009-12964-C05-03 (MICINN, Spain), and FIS2012-38866-C05-01 (MEC, Spain).


\begin{thebibliography}{99}

\bibitem{eq_review}  Wilson K G and  Kogut J 1974 {\sl Phys. Rep.} {\bf 12} 75

\bibitem{marro:book}  Marro J and Dickman R 1999 {\em Nonequilibrium Phase Transitions in Lattice Models} (Cambridge University Press, Cambridge, UK)

\bibitem{odor:2004} \'Odor G 2004 {\sl Rev. Mod. Phys.} {\bf 76} 663

\bibitem{munoz:2011}  Mu\~noz M A 2011 {\sl AIP Conf. Proc.} {\bf 1332} 111

\bibitem{meyers:2011} {\em Complex Systems in Finance and Econometrics}, edited by R. A. Meyers (Springer, New York, 2011).

\bibitem{sethna:2001}  Sethna J P,  Dahmen K A and  Myers C R 2001 {\sl Nature (London)}  {\bf 419} 242

\bibitem{albert:2002}  Albert R and  Barab{\'a}si A--L 2002 {\sl Rev. Mod. Phys.} {\bf 74} 47

\bibitem{belitz:2005}  Belitz D,  Kirkpatrick T R and  Vojta T 2005 {\sl Rev. Mod. Phys.} {\bf 77} 579

\bibitem{grinstein:1995} Grinstein G 1995 in {\em Scale Invariance, Interfaces, and Non-Equilibrium Dynamics} edited by  McKane A J,  Droz M,  Vannimenus J and Wolf
D (Plenum Press, New York)

\bibitem{schmittmann:2000}  Schmittmann B and  Zia R K P 2000 in {\em Phase transitions and critical phenomena} {\bf 17} edited by C. Domb and J. L. Lebowitz (Academic Press, London)

\bibitem{krug:1997}  Krug J 1997 {\sl Adv. Phys.} {\bf 46} 139

\bibitem{kardar:1986}  Kardar M,  Parisi G and  Zhang Y--C 1986 {\sl Phys. Rev. Lett.} {\bf 56} 889

\bibitem{sasamoto:2010}  Sasamoto T and  Spohn H 2010 {\sl Phys. Rev. Lett.} {\bf 104} 230602

\bibitem{amir:2011}  Amir G,  Corwin I and  Quastel J 2011  {\sl Commun. Pure Appl. Math.} {\bf 64} 466

\bibitem{calabrese:2011}  Calabrese P and  Le Doussal P 2011 {\sl Phys. Rev. Lett.} {\bf 106} 250603

\bibitem{praehofer:2000}  Pr\"ahofer M and  Spohn H 2000 {\sl Phys. Rev. Lett.} {\bf 84} 4882;
Pr\"ahofer M and  Spohn H 2000 {\sl Physica A} {\bf 279} 342

\bibitem{kriecherbauer:2010}  Kriecherbauer T and  Krug J 2010 {\sl J. Phys. A: Math. Theor.} {\bf 43} 403001

\bibitem{corwin:2012} Corwin I 2012 {\sl Random Matrices: Theor. Appl.} {\bf 1} 1130001

\bibitem{alves:2011_etal}  Alves S G,  Oliveira T J and  Ferreira S C 2011  {\sl Europhys. Lett.} {\bf 96} 48003;
 Oliveira T J,  Ferreira S C and  Alves S G 2012 {\sl Phys. Rev. E} {\bf 85} 010601(R)

\bibitem{takeuchi_et_al}  Takeuchi K A and  Sano M 2010 {\sl Phys. Rev. Lett.} {\bf 104} 230601;
Takeuchi K A, Sano M,  Sasamoto T and  Spohn H 2011 {\sl Sci. Rep.} {\bf 1} 34

\bibitem{yunker:2013}
Yunker P J, Lohr M A, Still T,  Borodin  A,  Durian D J and Yodh A G 2013 {\sl Phys. Rev. Lett.} {\bf 110} 035501

\bibitem{kim:2009} K.-J. Kim {\em et al.},
 Kim K--J,  Lee J--C,  Ahn S--M,  Lee K--S,  Lee C--W,  Cho Y J,  Seo S,  Shin K--H,  Choe S--B and  Lee H--W
2009  {\sl Nature (London)} {\bf 458} 740

\bibitem{halpin-healy:2012}  Halpin--Healy T 2012 {\sl Phys. Rev. Lett.} {\bf 109} 170602

\bibitem{oliveira:2013}  Oliveira T J,  Alves S G and  Ferreira S C 2013 {\sl Phys. Rev. E} {\bf 87} 040102(R)

\bibitem{kelling:2011}  Kelling J and  \'Odor G 2011 {\sl Phys. Rev. E} {\bf 84} 061150

\bibitem{eklund:1991_et_al} Eklund E A,  Bruinsma R, Rudnick J and  Stanley Williams R 1991 {\sl Phys. Rev. Lett.} {\bf 67} 1759;
Paniago R, Forrest R, Chow P C, Moss S C, Parkin S S P and Cookson D 1997 {\sl Phys. Rev. B} {\bf 56} 13442;
Ojeda F, Cuerno R,  V\'azquez L and  Salvarezza R 2000 {\sl Phys. Rev. Lett.} {\bf 84} 3125

\bibitem{cuerno:2007} Cuerno R,  Castro M,  Mu\~noz--Garc\'ia J,  Gago R and  V\'azquez L 2007
{\sl Eur. J. Phys. Special Topics} {\bf 146} 427

\bibitem{dassarma:2002}  Das Sarma S,  Punyindu Chatraphorn P and  Toroczkai Z 2002 {\sl Phys. Rev. E} {\bf 65} 036144

\bibitem{ryu:2007}  Ryu K--S,  Akinaga H and  Shin S--C 2007 {\sl Nature Phys.} {\bf 3} 547

\bibitem{nicoli:2009} Nicoli M,  Cuerno R and Castro M  2009 {\sl Phys. Rev. Lett.} {\bf 102} 256102

\bibitem{misbah:2010}  Misbah C,  Pierre--Louis O and  Saito Y 2010  {\sl Rev. Mod. Phys.} {\bf 82} 981

\bibitem{zhao:1999}  Zhao Y--P, Drotar J T,  Wang G--C and Lu T--M 1999 {\sl Phys. Rev. Lett.} {\bf 82}, 4882

\bibitem{nicoli:2009b} Nicoli M,  Castro M  and  Cuerno R 2009  {\sl J. Stat. Mech.: Theor. Exp.}  P02036

\bibitem{castro:2012} Castro M, Cuerno R, Nicoli M, V\'azquez L and  Buijnsters J G 2012 {\sl New J. Phys.} {\bf 14} 103039

\bibitem{nicoli:2011} Nicoli M,  Cuerno R and  Castro M 2011 {\sl J. Stat. Mech.: Theor. Exp.}  P10030

\bibitem{kassner:2002}  Kassner K and  Misbah C  2002 {\sl Phys. Rev. E} {\bf 66} 026102


\bibitem{mukamel:2010} Mukamel D 2010 in {\em Long-Range Interacting Systems}, edited by  Dauxois T,  Ruffo S and  Cugliandolo L F (Oxford University Press, Oxford)

\bibitem{katzav:2003} E. Katzav  2003 {\sl Phys. Rev. E} {\bf 68} 031607


\bibitem{edelman:2008}  Edelman A and  Persson P--O  2005 {\em Numerical Methods for Eigenvalue Distributions of
Random Matrices} arXiv:math-ph/0501068

\bibitem{bornemann:2010}  Bornemann F 2010 {\sl Math. Comput.} {\bf 79} 871

\bibitem{krug:1991_et_al}  Krug J and  Spohn H 1991 in {\em Solids far from equilibrium}, edited by C. Godr\`eche (Cambridge University Press, Cambridge, England);
Blair-Stahn N D  2010 arXiv:math/1005.0649v1

\bibitem{nicoli:2010} Nicoli M,  Vivo E and  Cuerno R 2010 {\sl Phys. Rev. E} {\bf 82} 045202(R)

\bibitem{newman:1997}  Newman T J and  Swift M R  1997 {\sl Phys. Rev. Lett.} {\bf 79} 2261

\bibitem{canet:2010} Canet L,  Chate H, Delamotte B and Wschebor  N 2010
{\sl Phys. Rev. Lett.} {\bf 104} 150601

\bibitem{batchelor:book}  Batchelor G K  2000 {\em An Introduction to Fluid Dynamics} (Cambridge University Press, Cambridge, UK)

\bibitem{hoyle:book} Hoyle R 2006 {\em Pattern formation: an introduction to methods} (Cambridge University Press, Cambridge, UK)

\bibitem{castro:2007}
Castro M,  Mu\~noz--Garc\'ia,  Cuerno R,  Garc\'ia Hern\'andez M and  V\'azquez L 2007 {\sl New J. Phys.} {\bf 9} 102



\end{thebibliography}
\end{document}